\documentclass[english]{article}
\usepackage[T1]{fontenc}
\usepackage[utf8]{inputenc}
\usepackage[letterpaper]{geometry}
\usepackage{amsmath}
\usepackage{amssymb}

\makeatletter
\newcommand{\lyxaddress}[1]{
\par {\raggedright #1
\vspace{1.4em}
\noindent\par}
}

\usepackage{mcite}

\makeatother

\usepackage{babel}

\begin{document}

\title{Quantization of the Damped Harmonic Oscillator Revisited}

\author{M.C. Baldiotti%
\thanks{baldiott@fma.if.usp.br%
}, R. Fresneda%
\thanks{fresneda@gmail.com%
}, and D.M. Gitman%
\thanks{gitman@dfn.if.usp.br%
}}

\maketitle

\lyxaddress{Instituto de Física, Universidade de São Paulo,\\
Caixa Postal 66318-CEP, 05315-970 São Paulo, S.P., Brasil}
\begin{abstract}
We return to the description of the damped harmonic oscillator by
means of a closed quantum theory with a general assessment of previous
works, in particular the Bateman-Caldirola-Kanai model and a new model
recently proposed by one of the authors. We show the local equivalence
between the two models and argue that latter has better high energy
behavior and is naturally connected to existing open-quantum-systems
approaches.
\end{abstract}

\section{Introduction}

The problem of constructing a quantum theory for the damped harmonic
oscillator as well as for similar dissipative systems, e.g. radiating
point-like charge, has attracted attention for already more than 50
years. In spite of the success of interaction-with-reservoir approaches,
we feel there is still room for some formal developments in the direction
of a closed theory approach. In this article we analyze a novel Lagrangian
model for the damped harmonic oscillator, which was recently proposed
in \cite{Gitman2007kg} as a particular case of a general procedure
for finding action functionals for non-Lagrangian equations of motion.
Even though we find it is locally equivalent to the renowned Bateman-Caldirola-Kanai
(BCK) model \cite{Bateman1931,Caldirola1941,Kanai1950} (see Section
\ref{sec:Bateman-Caldirola-Kanai-Theory} for a revision), a complete
global equivalence is absent, and we believe it has some formal advantages
over its predecessor, regarding the high energy behavior of solutions
to the Schrödinger equation.  Notwithstanding these formal discrepancies
between the two models, we show that they share a very close physical
interpretation with regard to their asymptotic behavior in time and
to their physical functions. 

Ever since the proposition of the BCK model, there have been divided
opinions as to whether it really describes the damped harmonic oscillator
or dissipation. There are those who dispute it as a possible dissipative
model \cite{Senitzky1960,Ray1979,Greenberger1979}, and those, in
addition to the original authors, who maintain it accounts for some
form of dissipation \cite{Dodonov1978,Dodonov1979}. As we explain
further on, the BCK theory is not defined globally in time, and many
of the pathologies usually appointed to the quantum theory can be
seen as an artifact of its infinite time indefinition. Nonetheless,
regardless of these disputes, there has been fruitful applications
of the BCK theory as a model of dissipation, at least in the case
of the  canonical description of the Fabry-Pérot cavity \cite{Colegrave1981}.

The underlying understanding of dissipative systems is that they are
physically part of a larger system, and dissipation is a result of
a non-elastic interaction between the reservoir and the subsystem.
Thus, quantization can conceivably follow two different approaches:
the first one takes the classical equations of motion of the system
and applies to them formal quantization methods, trying to overcome
the difficulties related to the ``non-Lagrangian'' nature of the system.
Without attempting to exhaust all the literature on this matter, we
cite, for instance, the use of canonical quantization of classical
actions \cite{Hasse1975,Okubo1981}, Fermi quantization \cite{Latimer2005},
path-integral quantization \cite{Kochan2010}, doubling of the degrees
of freedom \cite{BLASONE2004,CELEGHINI1992}, group theory methods
\cite{Villarroel1984}, complex classical coordinates \cite{Dekker1977},
propagator methods \cite{Tikochinsky1978}, non-linear Schrödinger
equation \cite{kostin1972}, and finally a constrained dynamics approach
\cite{Gitman2007b}. The second approach aims at constructing a quantum
theory of the subsystem by averaging over the reservoir, see e.g.
\cite{Senitzky1960,*Feynman1963,*Haken1975,*Caldeira:1982iu,*Pedrosa1984,*Walls1985,*Grabert1988,*Yu1994,*Joichi1997}.
It is immediately clear that two different outcomes must follow from
these approaches. Indeed, in the first case, we consider the system
as being closed, and thus naturally only pure states can result, and
they represent physical states without any further restrictions. In
the second case, the subsystem's states will be necessarily described
by a statistical operator, such that only mixed states are physically
sensible. In spite of the general understanding that the second approach
seems to be more physical and probably should produce a more adequate
quantum theory for dissipative systems, in this article we wish once
again to analyze the first, formal, approach. Its comparison to the
second one will be given in a future work. Our analysis will be particular
to the damped harmonic oscillator as an example of a dissipative system,
and it will be devoted to the comparison of two different canonical
quantizations.

In Section 2 we revise the BCK theory, emphasizing problems which
were not considered before. In Section 3 we present the first-order
theory, define some useful physical quantities, and we also construct
the coherent and squeezed states of the first-order theory in order
to obtain the classical limit. In Section 4, we prove the local equivalence
of the theories presented in Sections 2 and 3, and discuss possible
divergences related to their global behavior. In Section 5 we present
some final remarks and discuss our results as well as open problems.

\section{BCK Theory\label{sec:Bateman-Caldirola-Kanai-Theory}}

One of the peculiarities of dissipative systems, which hindered early
quantization attempts, was the non-Lagrangian nature of the classical
equations of motion. In the particular case of the damped harmonic
oscillator of constant frequency $\omega$ and friction coefficient
$\alpha>0$, the second-order equation of motion\begin{equation}
\ddot{q}+2\alpha\dot{q}+\omega^{2}q=0\label{eq:eqm-dho}\end{equation}
cannot be directly obtained as the Euler-Lagrange (EL) equation of
any Lagrangian, since it fails to satisfy the Helmholtz conditions
\cite{Helmholtz1887}. Nevertheless, there is an equivalent second-order
equation for which a variational principle can be found, namely,\begin{equation}
e^{2\alpha t}\left(\ddot{q}+2\alpha\dot{q}+\omega^{2}q\right)=0\,.\label{eq:equiv-dho-eqm}\end{equation}
The exponential factor is known as the integrating multiplier, and
it is enough to make the above equation satisfy the Helmholtz conditions
\cite{Starlet1982}. The fact that a Lagrangian can always be found
for the one-dimensional problem such that its EL equation is equivalent
to a given second-order equation was established by Darboux \cite{Darboux1894}.

As was already mentioned, the equation (\ref{eq:eqm-dho}) is traditionally
considered to be non-Lagrangian, albeit the existence of a questionable
action functional \cite{Ray1979,Greenberger1979} which reproduces
the equivalent equations of motion (\ref{eq:equiv-dho-eqm}). In this
respect, we have to mention that an action principle for the equation
of motion (\ref{eq:equiv-dho-eqm}) was first proposed by Bateman
\cite{Bateman1931} in terms of the Lagrangian \begin{equation}
L_{B}=\frac{1}{2}\left(\dot{q}^{2}-\omega^{2}q^{2}\right)e^{2\alpha t}\,.\label{eq:caldirola-kanai}\end{equation}
If Bateman had constructed the corresponding Hamiltonian formulation,
he would have discovered that the corresponding Hamiltonian theory
is canonical without constraints and with Hamiltonian\begin{equation}
H_{BCK}\left(q,p\right)=\frac{1}{2}\left[e^{-2\alpha t}p^{2}+\omega^{2}e^{2\alpha t}q^{2}\right]\,.\label{eq:caldirola-kanai-hamiltonian}\end{equation}
This Hamiltonian was proposed independently by Caldirola and Kanai
\cite{Caldirola1941,Kanai1950} to describe the damped harmonic oscillator
in the framework of quantum mechanics. Consequently, we write the
subscript BCK (Bateman-Caldirola-Kanai) to label the Hamiltonian.

Formal canonical quantization of the Lagrangian action (\ref{eq:caldirola-kanai})
is straightforward,\begin{equation}
\left[\hat{q},\hat{p}\right]=i\,,\,\,\left[\hat{q},\hat{q}\right]=\left[\hat{p},\hat{p}\right]=0,\ \hat{H}_{BCK}=H_{BCK}\left(\hat{q},\hat{p}\right)\,,\label{eq:BCK-quantum-theory}\end{equation}
and coincides with Caldirola and Kanai's quantum theory. Solutions
to the Schrödinger equation with Hamiltonian $\hat{H}_{BCK}$ have
been found in the form \begin{equation}
\psi_{n}^{BCK}\left(q,t\right)=\left(2^{n}n!\right)^{-1/2}\left(\frac{\tilde{\omega}}{\pi}\right)^{1/4}\exp\left(-iE_{n}t+\frac{\alpha t}{2}-\left(\tilde{\omega}+i\alpha\right)\frac{q^{2}}{2}e^{2\alpha t}\right)H_{n}\left(\sqrt{\tilde{\omega}}qe^{\alpha t}\right)\,,\label{eq:pseudostationary-state}\end{equation}
where $E_{n}=\tilde{\omega}\left(n+1/2\right)$, $\tilde{\omega}=\sqrt{\omega^{2}-\alpha^{2}}$
and $H_{n}$ are Hermite polynomials. These are the familiar pseudostationary
states \cite{Kerner1958,Bopp.F.1962,Hasse1975} or loss-energy states
\cite{Dodonov1978} which are also eigenstates of the Hamiltonian
$\hat{H}_{BCK}+\frac{\alpha}{2}\left(\hat{q}\hat{p}+\hat{p}\hat{q}\right)$
with eigenvalues $E_{n}$. Even though $\left\vert \psi_{n}^{BCK}\right\vert ^{2}$
depends on time, the total probability $\int dq\left\vert \psi_{n}^{BCK}\right\vert ^{2}=1$
is time-independent, as can be seen by the transformation of variables
$q\mapsto q^{\prime}=qe^{\alpha t}$. Moreover, the mean value of
the BCK Hamiltonian in the pseudostationary states is constant, \begin{equation}
\left\langle \psi_{n}^{BCK}\right\vert \hat{H}_{BCK}\left\vert \psi_{n}^{CK}\right\rangle =\left\langle \psi_{n}^{BCK}\right\vert E_{n}-\frac{\alpha}{2}\left(\hat{q}\hat{p}+\hat{p}\hat{q}\right)\left\vert \psi_{n}^{BCK}\right\rangle =\frac{\omega^{2}}{\tilde{\omega}}\left(n+\frac{1}{2}\right)\,,\label{eq:average-ck-hamiltonian}\end{equation}
which is a reflection of the fact that in the classical theory defined
by (\ref{eq:caldirola-kanai}) the average of the Hamiltonian over
the period of one oscillation is constant. On the other hand, mean
values of the mechanical energy $E=\frac{1}{2}\left(\dot{q}^{2}+\omega^{2}q^{2}\right)$
decay exponentially with time, $\left\langle E\right\rangle _{n}=e^{-2\alpha t}\left\langle H_{BCK}\right\rangle _{n}$
\cite{Kerner1958,Hasse1975}. Coherent states for the BCK theory are
given in \cite{Dodonov1979}, for which the uncertainty relations
are\begin{equation}
\Delta q\Delta p=\frac{\omega}{2\tilde{\omega}}\geq\frac{1}{2}\,.\label{eq:manko-uncertainty}\end{equation}

Now we draw attention to the high-energy behavior of pseudostationary
states. In the appendix we consider asymptotic (in $n$) pseudostationary
functions, and for these one has the limiting eigenvalue equation\begin{equation}
\hat{H}_{BCK}\psi_{n}^{BCK}=\left[E_{n}-\frac{\alpha}{2}\left(\hat{q}\hat{p}+\hat{p}\hat{q}\right)\right]\psi_{n}^{BCK}=\left(E_{n}+\frac{i\alpha}{2}\right)\psi_{n}^{BCK}+O\left(n^{-1/4}\right)\,.\label{eq:limit-eigenequation}\end{equation}
The appearance of imaginary eigenvalues is actually explained by taking
into account the domain of the operator $\hat{q}\hat{p}+\hat{p}\hat{q}$,
which, as is shown in the appendix, does not include the asymptotic
part of $\psi_{n}^{BCK}$. In this connection, it should be noted
that the above is in contrast to \cite{Dodonov1978,Dodonov1979a},
where it is claimed that the pseudostationary states are eigenstates
of $\hat{H}_{BCK}$ with eigenvalues $\left(\tilde{\omega}+i\alpha\right)\left(n+1/2\right)$.

One overlooked aspect of the BCK theory is that EL equation (\ref{eq:equiv-dho-eqm})
obtained from the Lagrangian (\ref{eq:caldirola-kanai}) is only equivalent
to the equation of motion of the damped harmonic oscillator (\ref{eq:eqm-dho})
for finite times. The equation (\ref{eq:equiv-dho-eqm}) is indicative
that the theory described by (\ref{eq:caldirola-kanai}) is not globally
defined, i.e., it is not defined for infinite times. In effect, the
manifold difficulties which appear in connection to the $t\rightarrow\infty$
limit, such as the violation of the Heisenberg uncertainty principle
and the vanishing of the ground state energy for infinite times, can
be seen as consequence of inadvertently assuming equations (\ref{eq:eqm-dho})
and (\ref{eq:equiv-dho-eqm}) are equivalent for all values of the
time parameter.

\section{First-order action\label{sec:First-order-action}}

\subsection{Action, hamiltonization, and quantization}

Next we consider the canonical quantization of the damped harmonic
oscillator based on the alternative action proposed in \cite{Gitman2007kg}.
The idea was to reduce the second-order equations (\ref{eq:eqm-dho})
to the first-order system\begin{equation}
\dot{x}=y,\ \dot{y}=-\omega^{2}x-2\alpha y\,,\label{eq:1st-order-eqm-dho}\end{equation}
 for which, according to the general theory \cite{Gitman2007a,Gitman2006},
the action functional has the form \begin{equation}
S=\frac{1}{2}\int dt\left[y\dot{x}-x\dot{y}-\left(y^{2}+2\alpha xy+\omega^{2}x^{2}\right)\right]e^{2\alpha t}\,.\label{eq:gitman-kupriyanov}\end{equation}
The EL equations of motion derived from (\ref{eq:gitman-kupriyanov})
are locally equivalent to (\ref{eq:1st-order-eqm-dho}),\[
\frac{\delta S}{\delta x}=\left(\dot{y}+2\alpha y+\omega^{2}x\right)e^{2\alpha t}\,,\,\,\frac{\delta S}{\delta y}=\left(\dot{x}-y\right)e^{2\alpha t}\,.\]
Note that, as in the case of the BCK theory, the theory fails to describe
the damped harmonic oscillator as the time approaches infinity, and
thus one should expect problems in the quantum theory such as violation
of the uncertainty principle for infinite times, for instance.

This action describes a singular system with second-class constraints,
and furthermore, these constraints are time-dependent (we follow the
terminology of the book \cite{Gitman:1990qh}). Even though the constraints
are explicitly time-dependent, it is still possible to write the Hamiltonian
formalism with the help of Dirac brackets and perform the canonical
quantization, as is explained in \cite{Gitman:1990qh}.

In order to do this, one must extend the initial phase space of canonical
variables $\eta=\left(x,y,p_{x},p_{y}\right)$ by the inclusion of
the time $t$ and its associated momentum $\varepsilon$. As a result,
the Poisson brackets between functions defined on the extended phase
space is\[
\left\{ F,G\right\} =\left(\frac{\partial F}{\partial x}\frac{\partial G}{\partial p_{x}}+\frac{\partial F}{\partial y}\frac{\partial G}{\partial p_{y}}+\frac{\partial F}{\partial t}\frac{\partial G}{\partial\varepsilon}\right)-F\leftrightarrow G\,.\]
With the above definition, the first-order equations of motion (\ref{eq:1st-order-eqm-dho})
are equivalently written in terms of Dirac brackets, \begin{equation}
\dot{\eta}=\left\{ \eta,H\left(x,y,t\right)+\epsilon\right\} _{D\left(\phi\right)}\,,\,\,\phi_{x}=\phi_{y}=0\,,\label{eq:constrained-eqm}\end{equation}
where the Hamiltonian $H\left(x,y,t\right)$ and the constraints $\phi$
are \begin{align}
 & H\left(x,y,t\right)=\frac{1}{2}\left(y^{2}+2\alpha xy+\omega^{2}x^{2}\right)e^{2\alpha t}\,,\label{eq:kg-hamiltonian}\\
 & \phi_{x}=p_{x}-\frac{1}{2}ye^{2\alpha t}\,,\,\,\phi_{y}=p_{y}+\frac{1}{2}xe^{2\alpha t}\,.\nonumber \end{align}
The nonzero commutation relations between the independent variables
are\begin{equation}
\left\{ x,y\right\} _{D\left(\phi\right)}=e^{-2\alpha t}\,.\label{eq:classical-commutators}\end{equation}
Quantization of this system follows the general method described in
\cite{Gitman:1990qh} for theories with time-dependent constraints.
One introduces time-dependent operators $\hat{\eta}\left(t\right)$
which satisfy the differential equations $d\hat{\eta}/dt=\left.i\left\{ \eta,\epsilon\right\} _{D\left(\phi\right)}\right\vert _{\eta=\hat{\eta}}$
with initial conditions subject to an analog of the Dirac quantization,
\begin{align*}
 & \left[\hat{\eta}\left(0\right),\hat{\eta}\left(0\right)\right]=\left.i\left\{ \eta,\eta\right\} _{D\left(\phi_{0}\right)}\right\vert _{\eta=\hat{\eta}\left(0\right)}\,,\\
 & \hat{p}_{x}\left(0\right)-\frac{1}{2}\hat{y}\left(0\right)=\hat{p}_{y}\left(0\right)+\frac{1}{2}\hat{x}\left(0\right)=0\,.\end{align*}
The above operatorial constraints allows us to work only in terms
of the independent operators $\hat{x}$ and $\hat{y}$. Let us define
$\hat{x}\left(0\right)\equiv\hat{q}$ and $\hat{y}\left(0\right)\equiv\hat{p}$,
so that the above quantum brackets have the familiar form \begin{equation}
\left[\hat{q},\hat{p}\right]=i\,,\,\,\left[\hat{q},\hat{q}\right]=\left[\hat{p},\hat{p}\right]=0,\ \ \hat{x}\left(0\right)\equiv\hat{q},\ \ \hat{y}\left(0\right)\equiv\hat{p}\,.\label{eq:qp-operators}\end{equation}
The differential equations for $\hat{x}$ and $\hat{y}$ can be easily
integrated to\begin{align}
 & \hat{x}\left(t\right)=e^{-\alpha t}\hat{q}\,,\,\,\hat{x}\left(0\right)\equiv\hat{q}\,,\,\,\hat{y}\left(t\right)=e^{-\alpha t}\hat{p}\,,\,\,\hat{y}\left(0\right)\equiv\hat{p}\,.\label{eq:operator-solutions}\end{align}
The quantum Hamiltonian is obtained from the classical Hamiltonian
(\ref{eq:kg-hamiltonian}) as a function of the operators $\hat{x}\left(t\right)$
and $\hat{y}\left(t\right)$ (\ref{eq:operator-solutions}): it does
not depend on time at all,

\begin{equation}
\hat{H}=H\left(\hat{x}\left(t\right),\hat{y}\left(t\right),t\right)=\frac{1}{2}\left[\hat{p}^{2}+\alpha\left(\hat{q}\hat{p}+\hat{p}\hat{q}\right)+\omega^{2}\hat{q}^{2}\right]\,,\label{eq:kg-quantum-hamiltonian}\end{equation}
where we have used Weyl (symmetric) ordering for the mixed product
$2\alpha xy$. The Hamiltonian $\hat{H}$ governs the time-evolution
of the state vector in the Schrödinger picture, and it has appeared
in a number of different contexts: in \cite{Stevens1958,Colegrave1981}
with regard to the electromagnetic field in a resonant cavity; in
the quantization of the complex symplectic theory \cite{Dekker1981};
and in \cite{Isar1999} in connection to the Lindblad theory of open
quantum systems for the damped harmonic oscillator.

Since the Hamiltonian is time-independent, the evolution operator
is given simply by $U\left(t\right)=e^{-i\hat{H}t}$, with $\hat{H}$
given by (\ref{eq:kg-quantum-hamiltonian}). The Heisenberg operators
$\check{x}$ and $\check{y}$ corresponding to the classical variables
$x$ and $y$ are\begin{align*}
 & \check{x}=U^{-1}\hat{x}\left(t\right)U=e^{-\alpha t}\left(\cos\tilde{\omega}t+\frac{\alpha}{\tilde{\omega}}\sin\tilde{\omega}t\right)\hat{q}+\frac{1}{\tilde{\omega}}e^{-\alpha t}\sin\left(\tilde{\omega}t\right)\hat{p}\,,\\
 & \check{y}=U^{-1}\hat{y}\left(t\right)U=e^{-\alpha t}\left(\cos\tilde{\omega}t-\frac{\alpha}{\tilde{\omega}}\sin\tilde{\omega}t\right)\hat{p}-\frac{\omega^{2}}{\tilde{\omega}}e^{-\alpha t}\sin\left(\tilde{\omega}t\right)\hat{q}\,,\\
 & \check{x}\left(0\right)\equiv\hat{q}\,,\,\,\check{y}\left(0\right)\equiv\hat{p}\,,\,\,\left[\hat{q},\hat{p}\right]=i\,,\,\,\left[\hat{q},\hat{q}\right]=\left[\hat{p},\hat{p}\right]=0\,.\end{align*}
From the above expressions, one also finds\[
\frac{d\check{x}}{dt}=\check{y}\,,\,\frac{d\check{y}}{dt}=-\omega^{2}\check{x}-2\alpha\check{y}\,,\]
which coincide in form with the classical equations,\begin{equation}
\frac{d\check{x}}{dt}=\left.\left\{ x,H+\epsilon\right\} _{D\left(\phi\right)}\right\vert _{\eta=\check{\eta}}\,,\,\,\frac{d\check{y}}{dt}=\left.\left\{ y,H+\epsilon\right\} _{D\left(\phi\right)}\right\vert _{\eta=\check{\eta}}\,.\label{eq:heisenberg-equations}\end{equation}
Thus, Heisenberg equations (\ref{eq:heisenberg-equations}) reproduce
the classical equations of motion, and therefore mean values of $x$
and $y$ follow classical trajectories.

Moreover, the only nonzero commutator becomes\[
\left[\check{x},\check{y}\right]=ie^{-2\alpha t}\,,\]
which matches the classical Dirac bracket (\ref{eq:classical-commutators}).
Thus, the resulting quantum theory at least obeys the correspondence
principle.

One can easily find solutions of the Schrödinger equation by making
a (time-independent) unitary transformation $\hat{H}_{\tilde{\omega}}=\hat{S}^{-1}\hat{H}\hat{S},$
\begin{equation}
\hat{S}=\exp\left(-\frac{i\alpha}{2}\hat{q}^{2}\right)\,,\,\,\hat{S}^{-1}\hat{p}\hat{S}=\hat{p}-\alpha\hat{q}\,,\label{eq:s-transform}\end{equation}
one obtains \begin{equation}
\hat{H}_{\tilde{\omega}}=\frac{1}{2}\left(\hat{p}^{2}+\tilde{\omega}^{2}\hat{q}^{2}\right),\ \ \tilde{\omega}=\sqrt{\omega^{2}-\alpha^{2}}\,.\label{eq:undamped-hamiltonian}\end{equation}
The Hamiltonian (\ref{eq:undamped-hamiltonian}) has the familiar
stationary states \begin{align*}
 & \psi_{n}^{\tilde{\omega}}\left(t,q\right)=\exp\left(-itE_{n}\right)\psi_{n}^{\tilde{\omega}}\left(q\right)\,,\\
 & \psi_{n}^{\tilde{\omega}}\left(q\right)=\left(2^{n}n!\right)^{-1/2}\left(\frac{\tilde{\omega}}{\pi}\right)^{1/4}\exp\left(-\frac{\tilde{\omega}}{2}q^{2}\right)H_{n}\left(\sqrt{\tilde{\omega}}q\right)\,,\\
 & \hat{H}_{\tilde{\omega}}\psi_{n}^{\tilde{\omega}}=E_{n}\psi_{n}^{\tilde{\omega}}\,,\,\, E_{n}=\tilde{\omega}\left(n+\frac{1}{2}\right)\,.\end{align*}
Therefore, the wave functions\begin{equation}
\psi_{n}\left(q\right)=\hat{S}\psi_{n}^{\tilde{\omega}}\left(q\right)=\left(2^{n}n!\right)^{-1/2}\left(\frac{\tilde{\omega}}{\pi}\right)^{1/4}\exp\left(-\left(\tilde{\omega}+i\alpha\right)\frac{q^{2}}{2}\right)H_{n}\left(\sqrt{\tilde{\omega}}q\right)\label{eq:GK-eigenfunctions}\end{equation}
are eigenfunctions of the Hamiltonian $\hat{H}$, and the solutions
to the corresponding Schrödinger equation are\begin{align*}
 & \psi_{n}\left(t,q\right)=\hat{S}\psi_{n}^{\tilde{\omega}}\left(t,q\right)=\exp\left(-itE_{n}\right)\psi_{n}\left(q\right)\,,\\
 & \hat{H}\psi_{n}\left(q\right)=E_{n}\psi\left(q\right)\,.\end{align*}
One can also check directly that (\ref{eq:GK-eigenfunctions}) are
eigenfunctions of $\hat{H}$ with eigenvalue $E_{n}$ by using properties
of Hermite functions.

We now define some useful quantities to be used in Subsection \ref{sub:Physical-Equivalence}
for the purpose of establishing the physical equivalence between the
approaches presented here. Let us write the classical Lagrangian energy
as

\[
\mathcal{E}_{L}\equiv\frac{\partial L}{\partial\dot{x}}\dot{x}+\frac{\partial L}{\partial\dot{y}}\dot{y}-L=\frac{1}{2}\left(y^{2}+2\alpha xy+\omega^{2}x^{2}\right)e^{2\alpha t}\,.\]
The corresponding Weyl-ordered Schrödinger operator for the Lagrangian
energy is\begin{equation}
\hat{\mathcal{E}}_{L}=\left.\mathcal{E}_{L}\left(\eta\right)\right\vert _{\eta=\hat{\eta}}=\frac{1}{2}\left[\hat{p}^{2}+\alpha\left(\hat{q}\hat{p}+\hat{p}\hat{q}\right)+\omega^{2}\hat{q}^{2}\right]=\hat{H}\,,\label{eq:kg-conserved-energy}\end{equation}
which coincides with the Hamiltonian and thus with its Heisenberg
representation, and is therefore conserved. Likewise, we define the
following conserved ``energy'' in the BCK approach, \begin{align*}
 & \mathcal{E}=\frac{1}{2}\left(\dot{q}^{2}+2\alpha q\dot{q}+\omega^{2}q^{2}\right)e^{2\alpha t}=H_{BCK}+\alpha qp\,,\\
 & \frac{d}{dt}\mathcal{E}=\left(\dot{q}+\alpha q\right)e^{2\alpha t}\frac{\delta S}{\delta q}\,,\end{align*}
which is constant on-shell. Its image as a Weyl-ordered Schrödinger
operator is\begin{equation}
\hat{\mathcal{E}}=\hat{H}_{BCK}+\frac{\alpha}{2}\left(\hat{q}\hat{p}+\hat{p}\hat{q}\right)\,,\,\,\hat{\mathcal{E}}\psi_{n}^{BCK}=E_{n}\psi_{n}^{BCK}\,.\label{eq:BCK-conserved-energy}\end{equation}

Now consider the mechanical energy in the theory with Lagrangian (\ref{eq:gitman-kupriyanov})
at $\alpha=0$:

\[
E_{M}=\frac{1}{2}\left(y^{2}+\omega^{2}x^{2}\right)\,.\]
The corresponding operator is equal to\begin{equation}
\hat{E}_{M}=\frac{1}{2}\left(\hat{p}^{2}+\omega^{2}\hat{q}^{2}\right)e^{-2\alpha t}=e^{-2\alpha t}\hat{H}-e^{-2\alpha t}\frac{\alpha}{2}\left(\hat{q}\hat{p}+\hat{p}\hat{q}\right)\,.\label{eq:1st-order-mech-energy}\end{equation}
The mean value of $\hat{E}_{M}$ in the energy eigenstates is\[
\left\langle \psi_{n}\right\vert \hat{E}_{M}\left\vert \psi_{n}\right\rangle =e^{-2\alpha t}\frac{\omega^{2}}{\tilde{\omega}}\left(n+\frac{1}{2}\right)\,.\]
Finally, we consider the observable defined by the mechanical energy
in the BCK description, \begin{equation}
E=\frac{1}{2}\left(\dot{q}^{2}+\omega^{2}q^{2}\right)=\frac{1}{2}\left(e^{-2\alpha t}p^{2}+\omega^{2}e^{2\alpha t}q^{2}\right)e^{-2\alpha t}=H_{BCK}e^{-2\alpha t}\,.\label{eq:bck-mech-energy}\end{equation}
Thus, by (\ref{eq:average-ck-hamiltonian}), \[
\left\langle \psi_{n}^{BCK}\right\vert \hat{E}\left\vert \psi_{n}^{BCK}\right\rangle =e^{-2\alpha t}\frac{\omega^{2}}{\tilde{\omega}}\left(n+\frac{1}{2}\right)=\left\langle \psi_{n}\right\vert \hat{E}_{M}\left\vert \psi_{n}\right\rangle \,.\]

\subsection{Semiclassical description}

\subsubsection{Coherent states}

Finally, we obtain semiclassical states for the damped harmonic oscillator
from the coherent states of the simple harmonic oscillator using the
unitary transformation $\hat{S}$ (\ref{eq:s-transform}). To this
end, we introduce first creation and annihilation operators $\hat{a}^{+}$
and $\hat{a}$ and the corresponding coherent states $\left\vert z\right\rangle $,\begin{align}
 & \hat{a}=\frac{1}{\sqrt{2\tilde{\omega}}}\left(\tilde{\omega}\hat{q}+i\hat{p}\right)\,,\,\,\hat{a}^{+}=\frac{1}{\sqrt{2\tilde{\omega}}}\left(\tilde{\omega}\hat{q}-i\hat{p}\right)\,,\,\,\left[\hat{a},\hat{a}^{+}\right]=1\,,\nonumber \\
 & \left\vert z\right\rangle =D\left(z\right)\left\vert 0\right\rangle \,,\,\, D\left(z\right)=\exp\left(z\hat{a}^{+}-\bar{z}\hat{a}\right)\,,\,\, a\left\vert z\right\rangle =z\left\vert z\right\rangle \,.\label{eq:coherent-states}\end{align}
In terms of these creation and annihilation operators, the Hamiltonian
(\ref{eq:undamped-hamiltonian}) is \[
\hat{H}_{\tilde{\omega}}=\tilde{\omega}\left(\hat{a}^{+}\hat{a}+\frac{1}{2}\right)\,.\]
Thus, the coherent states for the Hamiltonian $\hat{H}$ are $\hat{S}\left\vert z\right\rangle $
and the mean values of $\check{x}$ and $\check{y}$ in these coherent
states are\begin{align*}
 & \left\langle x\right\rangle \equiv\left\langle z\right\vert \hat{S}^{-1}\check{x}\hat{S}\left\vert z\right\rangle =\frac{1}{\sqrt{2\tilde{\omega}}}e^{-\alpha t}\left(ze^{-i\tilde{\omega}t}+\bar{z}e^{i\tilde{\omega}t}\right)\,,\\
 & \left\langle y\right\rangle =i\sqrt{\frac{\tilde{\omega}}{2}}e^{-\alpha t}\left(\bar{z}e^{i\tilde{\omega}t}-ze^{-i\tilde{\omega}t}\right)-\alpha\left\langle x\right\rangle \,.\end{align*}
One can now easily verify that the mean values of the coordinates
$x$ and $y$ follow the classical trajectories,\[
\frac{d}{dt}\left\langle x\right\rangle =\left\langle y\right\rangle \,,\,\,\frac{d}{dt}\left\langle y\right\rangle =-\omega^{2}\left\langle x\right\rangle -2\alpha\left\langle y\right\rangle \,.\]
The pathological behavior of the first-order theory with regard to
the limit $t\rightarrow\infty$ can be seen here in the computation
of the uncertainty relation \begin{equation}
\Delta x\Delta y=\frac{1}{2}e^{-2\alpha t}\frac{\omega}{\tilde{\omega}}\,.\label{eq:time-dep-uncertainty}\end{equation}
The unphysical result of the above violation of the uncertainty principle
is an indication that the first-order theory is not defined for all
values of the time parameter, as was pointed out in connection to
the classical equations of motion and to the commutation relation
between $\hat{x}$ and $\hat{y}$. Another indication of the failure
of the theory at infinite times is in the observation that the radius
of the trajectory of the mean values vanishes:\[
\rho\left(z=re^{i\theta}\right)=\sqrt{\left\langle x\right\rangle ^{2}+\left\langle y\right\rangle ^{2}}=\sqrt{\frac{2}{\tilde{\omega}}}e^{-\alpha t}r\sqrt{1+\alpha^{2}\cos\left(\tilde{\omega}t-\theta\right)+\alpha\sin2\left(\tilde{\omega}t-\theta\right)}\,.\]

\subsubsection{Squeezed-state}

One can also consider the family of conserved creation and annihilation
operators\begin{align*}
 & \hat{b}\left(t\right)=\cosh\xi e^{i\tilde{\omega}t}\hat{a}+\sinh\xi e^{-i\tilde{\omega}t}\hat{a}^{\dagger}\,,\,\,\hat{b}^{\dagger}\left(t\right)=\cosh\xi e^{-i\tilde{\omega}t}\hat{a}^{\dagger}+\sinh\xi e^{i\tilde{\omega}t}\hat{a}\,,\\
 & \left[\hat{b}\left(t\right),\hat{b}^{\dagger}\left(t\right)\right]=1\,,\,\,\frac{d}{dt}\hat{b}=\frac{d}{dt}\hat{b}^{\dagger}=0\,,\end{align*}
and construct squeezed coherent states $\left\vert z,\xi\right\rangle =\exp\left(z\hat{b}^{\dagger}-\bar{z}\hat{b}\right)\left\vert 0\right\rangle $
\cite{Dodonov}. For $\xi=0$ one arrives at the previous coherent
states, \[
\left\langle z,0\right\vert \hat{x}\left\vert z,0\right\rangle =\left\langle z\right\vert \hat{S}^{-1}\check{x}\hat{S}\left\vert z\right\rangle \,,\,\,\left\langle z,0\right\vert \hat{y}\left\vert z,0\right\rangle =\left\langle z\right\vert \hat{S}^{-1}\check{y}\hat{S}\left\vert z\right\rangle \,.\]
For arbitrary values of $\xi$ and with $\hat{x}$ and $\hat{y}$
given by (\ref{eq:operator-solutions}), one has for the mean values
of coordinates\begin{align*}
\left\langle z,\xi\right\vert \hat{x}\left\vert z,\xi\right\rangle  & =\frac{e^{-\alpha t}}{\sqrt{2\tilde{\omega}}}\left[\left(e^{-i\tilde{\omega}t}z+e^{i\tilde{\omega}t}\bar{z}\right)\cosh\xi-\left(e^{i\tilde{\omega}t}z+e^{-i\tilde{\omega}t}\bar{z}\right)\sinh\xi\right]\,,\\
\left\langle z,\xi\right\vert \hat{y}\left\vert z,\xi\right\rangle  & =i\sqrt{\frac{\tilde{\omega}}{2}}e^{-\alpha t}\left[\left(e^{i\tilde{\omega}t}\bar{z}-e^{-i\tilde{\omega}t}z\right)\cosh\xi-\left(e^{i\tilde{\omega}t}z-e^{-i\tilde{\omega}t}\bar{z}\right)\sinh\xi\right]-\alpha\left\langle z,\xi\right\vert \hat{x}\left\vert z,\xi\right\rangle \,.\end{align*}
The main interest in squeezed states is that they allow one to change
the uncertainty in either direction $x$ or $y$ by adjusting the
parameter $\xi$. For example, the uncertainty in $x$ can be written
for arbitrary values of $\xi$ as \[
\left(\Delta x\right)^{2}=\frac{1}{2\tilde{\omega}}e^{-2\alpha t}\left[\left(\cosh\xi+\sinh\xi\right)^{2}-4\cosh\xi\sinh\xi\cos^{2}\tilde{\omega}t\right]\geqslant0\,,\]
and it reduces to the preceding coherent states calculations for $\xi=0$.

\section{Comparison of the BCK system and the first-order system}

\subsection{BCK model as a transformation of the first-order system}

Here we show how one can obtain the classical and quantum description
of the BCK damped harmonic oscillator as a canonical transformation
of the first-order approach (\ref{sec:First-order-action}). At the
classical level, both systems are transformed one into the other by
means of the following time-dependent canonical transformation%
\footnote{The generating function for these canonical transformations depending
on the new and old momenta is\[
F\left(p_{x},p_{y},p,\Omega_{2},t\right)=-2e^{-2\alpha t}\left(\Omega_{2}-p_{y}\right)p_{x}-2e^{-2\alpha t}pp_{y}+e^{-2\alpha t}p\Omega_{2}\,.\]
}\begin{align*}
 & q=\frac{1}{2}\left(x-2e^{-2\alpha t}p_{y}\right)\,,\,\, p=p_{x}+\frac{1}{2}e^{2\alpha t}y\,,\\
 & \Omega^{1}=\frac{1}{2}y-p_{x}e^{-2\alpha t}\,,\,\,\Omega_{2}=p_{y}+\frac{1}{2}xe^{2\alpha t}\,,\end{align*}
where $\left(q,p\right)$ and $\left(\Omega^{1},\Omega_{2}\right)$
are new pairs of canonical variables. In these new variables, the
equation of motion become Hamiltonian:\begin{align*}
 & \dot{q}=\left\{ q,H_{BCK}\right\} \,,\,\,\dot{p}=\left\{ p,H_{BCK}\right\} \,,\,\,\Omega=0\,,\\
 & H_{BCK}=\frac{1}{2}\left[e^{-2\alpha t}p^{2}+\omega^{2}e^{2\alpha t}q^{2}\right]\,,\end{align*}
where the Hamiltonian $H_{BCK}\left(q,p,t\right)$ is the canonically
transformed Hamiltonian $H\left(x,y,t\right)$ (\ref{eq:kg-hamiltonian})
on the equivalent constraint surface $\Omega=0$.

It is useful to write the following relation between old coordinates
$x$ and $y$ and the new variables $q$ and $p$:\begin{eqnarray}
x & = & q+O\left(\Omega\right)\,,\nonumber \\
y & = & e^{-2\alpha t}p+O\left(\Omega\right)\,.\label{eq:physical-relations}\end{eqnarray}
This relation shows that $x$ is physically equivalent to $q$, while
$y$ is physically equivalent to $e^{-2\alpha t}p$, since they coincide
on the constraint surface $\Omega=0$.

The quantum theory can be readily obtained by a quantum time-dependent
canonical transformation%
\footnote{A proof that $\hat{x}\hat{p}+\hat{p}\hat{x}$ is self-adjoint is provided
in the appendix.%
} \begin{equation}
\hat{D}=\exp\left[\frac{i\alpha t}{2}\left(\hat{q}\hat{p}+\hat{p}\hat{q}\right)\right]\,,\label{eq:time-dep-unitary-transf}\end{equation}
which is suggested from the classical generating function and the
relationship between old and new variables. The effect of $\hat{D}$
on the canonical variables is to make the dilation \[
\hat{D}^{-1}\hat{q}\hat{D}=e^{-\alpha t}\hat{q}\,,\,\,\hat{D}^{-1}\hat{p}\hat{D}=e^{\alpha t}\hat{p}\,.\]
The dilation operator $\hat{D}$ has been discussed in a more general
setting in \cite{Onofri1978} in connection with the description of
open systems by time-dependent Hamiltonians. There, as is the case
here, the dilation operator simplifies the calculation of the evolution
operator. 

The dilaton operator transforms the BCK Hamiltonian into the first-order
Hamiltonian $\hat{H}$,\[
\hat{H}=\hat{D}^{-1}\hat{H}_{BCK}\hat{D}-i\hat{D}^{-1}\frac{\partial\hat{D}}{\partial t}=\frac{1}{2}\left[\hat{p}^{2}+\alpha\left(\hat{q}\hat{p}+\hat{p}\hat{q}\right)+\omega^{2}\hat{q}^{2}\right]\,.\]
Since $\psi_{n}\left(q,t\right)$ satisfy the Schrödinger equation
with $\hat{H}$, it follows that $\hat{D}\psi_{n}\left(q,t\right)$
satisfy the Schrödinger equation with $\hat{H}_{BCK}$. The wave functions
$\hat{D}\psi_{n}\left(q,t\right)$ are indeed the pseudostationary
states, as can be seen by direct application of $\hat{D}$. Thus,
we write $\psi_{n}^{BCK}\left(q,t\right)\equiv\hat{D}\psi_{n}\left(q,t\right)$.
Similarly, it also follows that $U=\hat{D}\exp\left(-it\hat{H}\right)$
satisfies the Schrödinger equation with Hamiltonian $\hat{H}_{BCK}$,

\[
U\left(t\right)=\hat{D}\exp\left(-i\hat{H}t\right)\,,\,\, i\frac{\partial U}{\partial t}=\hat{H}_{BCK}U\,,\,\, U\left(0\right)=I\,.\]
One should not be to eager to jump to the conclusion that the two
theories here presented, the BCK theory and the first-order theory,
are physically equivalent solely on the grounds of the time-dependent
canonical transformation $\hat{D}$. The existence of this transformation
\emph{per se }is insufficient to prove physical equivalence, since
in principle, and at least locally, one can always construct a time-dependent
canonical transformation between any two given theories, classical
or quantum. For instance, given two quantum theories $T_{1}$ and
$T_{2}$, and corresponding evolution operators $U_{1}$ and $U_{2}$,
one can always write a general solution $\psi_{2}\left(t\right)$
of the Schrödinger equation of $T_{2}$ in terms of the general solution
$\psi_{1}\left(t\right)$ of $T_{1}$ by means of the transformation
$\psi_{2}\left(t\right)=U_{2}U_{1}^{-1}\psi_{1}\left(t\right)$. Besides
the necessary ingredient of the unitary transformation relating the
two theories, it is imperative to show that the physical observables
pertaining both theories are also unitarily equivalent in order to
prove physical equivalence. A proof which we postpone to the next
Section.

\subsection{Physical Equivalence\label{sub:Physical-Equivalence}}

In this Section we show that the two approaches presented in this
article are physically equivalent. In order to sum up the previous
results in a coherent whole, let us present both quantum theories
anew.

Starting with the BCK theory, it is defined by the Hamiltonian $\hat{H}_{BCK}$
(\ref{eq:BCK-quantum-theory}) written in terms of the canonically
conjugated operators $\hat{q}$ and $\hat{p}$ (\ref{eq:BCK-quantum-theory})
with the usual realization in terms of multiplication and derivation
operators in the Hilbert space $\mathcal{H}_{BCK}$ of square-integrable
states $\psi_{BCK}\left(q\right)$ with measure\[
\left\langle \psi_{BCK}\right.\left|\psi_{BCK}\right\rangle _{\mathcal{H}_{BCK}}=\int_{-\infty}^{+\infty}dq\bar{\psi}_{BCK}\left(q\right)\psi_{BCK}\left(q\right)\,.\]

The first-order theory is defined by the Hamiltonian $\hat{H}$ (\ref{eq:kg-quantum-hamiltonian})
and canonically conjugated operators $\hat{q}$ and $\hat{p}$ (\ref{eq:qp-operators})
realized as the usual multiplication and derivation operators in the
Hilbert space $\mathcal{H}$ of square-integrable states $\psi\left(q\right)$
with measure\[
\left\langle \psi\right.\left|\psi\right\rangle _{\mathcal{H}}=\int_{-\infty}^{+\infty}dq\bar{\psi}\left(q\right)\psi\left(q\right)\,.\]

We have seen that the time-dependent unitary transformation $\hat{D}$
(\ref{eq:time-dep-unitary-transf}) maps the two Hilbert spaces, \[
\hat{D}:\mathcal{H}\rightarrow\mathcal{H}_{BCK}\,,\,\,\psi\mapsto\psi_{BCK}=\hat{D}\psi\,,\]
and is a canonical transformation, \[
\hat{H}_{BCK}=\hat{D}\hat{H}\hat{D}^{-1}+i\frac{\partial\hat{D}}{\partial t}\hat{D}^{-1}\,.\]
Therefore, for every $\psi\left(t\right)$ solution of the Schrödinger
equation of of the first-order theory, $\hat{D}\psi\left(t\right)$
is a solution of the BCK Schrödinger equation. Furthermore, one has
$\left\langle \psi_{BCK}\right.\left|\psi_{BCK}\right\rangle _{\mathcal{H}_{BCK}}=\left\langle \psi\right.\left|\psi\right\rangle _{\mathcal{H}}$,
as can be seen by noting that $\hat{D}\psi\left(q\right)=e^{\alpha t/2}\psi\left(qe^{\alpha t}\right)$
and\[
\left\langle \psi_{BCK}\right.\left|\psi_{BCK}\right\rangle _{\mathcal{H}_{BCK}}=\int_{-\infty}^{+\infty}dqe^{\alpha t}\bar{\psi}\left(qe^{\alpha t}\right)\psi\left(qe^{\alpha t}\right)=\int_{-\infty}^{+\infty}dq\bar{\psi}\left(q\right)\psi\left(q\right)=\left\langle \psi\right.\left|\psi\right\rangle _{\mathcal{H}}\,.\]

Finally, to complete the proof of the physical equivalence, it remains
to show that any two physical observables $\hat{\mathcal{O}}_{BCK}$
and $\hat{\mathcal{O}}$ are $D$-equivalent, that is, $\hat{\mathcal{O}}_{BCK}=\hat{D}\hat{\mathcal{O}}\hat{D}^{-1}$.
One can check that this is indeed the case for the physical observables
previously considered, such as the mechanical energies (\ref{eq:1st-order-mech-energy},\ref{eq:bck-mech-energy})

\[
\hat{E}=\hat{D}\hat{E}_{M}\hat{D}^{-1}\,,\]
 and the conserved energies (\ref{eq:kg-conserved-energy},\ref{eq:BCK-conserved-energy}),\[
\hat{\mathcal{E}}=\hat{D}\hat{\mathcal{E}}_{L}\hat{D}^{-1}\,.\]

As a result of the equivalence, one can easily obtain the BCK coherent
states presented in \cite{Dodonov1979} by merely transforming the
coherent states (\ref{eq:coherent-states}) given in the context of
the first-order theory. Taking into account the relation $y=e^{-2\alpha t}p+\left\{ \Omega\right\} $
between the physical variables of the two theories (\ref{eq:physical-relations}),
one can see why the uncertainty relations we obtain decay exponentially
with time (\ref{eq:time-dep-uncertainty}), while those (\ref{eq:manko-uncertainty})
calculated in \cite{Dodonov1979} are constant. This has a simple
explanation in terms of the physical equivalence of $y$ and $e^{-2\alpha t}p$:
$y$ is physically equivalent to the physical momentum of the BCK
oscillator, and it is in terms of the physical momentum that the Heisenberg
uncertainty relations are violated (see Dekker \cite{Dekker1981}
for a comprehensive review).

\section{Conclusion}

In this article we have proved that the classical and quantum description
of the damped harmonic oscillator by the BCK time-dependent Hamiltonian
\cite{Bateman1931,Caldirola1941,Kanai1950} is locally equivalent
to the first order approach given in terms of a constrained system
\cite{Gitman2007b}. This equivalence allowed us to easily obtain
the evolution operator for the BCK oscillator and many other results
in a simpler manner, due to the time-independence of the quantum first-order
Hamiltonian. As has been pointed out (see Dekker \cite{Dekker1981}
for a comprehensive review), the BCK oscillator has a pathological
behavior for infinite times, since the (mechanical) energy mean values
and the Heisenberg uncertainty relation - between the coordinate and
the physical momentum - go to zero as time approaches infinity, so
that even the ground state's energy eventually vanishes. Despite these
shortcomings, the quantum theory has a well-defined classical limit,
and for time values less than $\frac{1}{2\alpha}\ln\frac{\omega}{\tilde{\omega}}$
the Heisenberg uncertainty principle is not violated. It is our understanding
that this unphysical behavior is a result of extending the proposed
theories beyond their validity. The fact that both theories are not
globally defined in time has its roots already at the classical level,
as a consequence of the non-Lagrangian nature of the equations of
motion of the damped harmonic oscillator. 

Finally, we recall the intriguing behavior of the asymptotic pseudostationary
states, which so far has escaped notice from all works dedicated to
the BCK oscillator. At first glance eq. (\ref{eq:limit-eigenequation})
implies that the BCK Hamiltonian loses self-adjointness as $n$ approaches
infinity. Fortunately, these asymptotic states are not in the domain
of the BCK Hamiltonian and thus pose no threat whatsoever. On the
other hand, there is no such constraint on the domain of the first-order
Hamiltonian $\hat{H}$, where there is no upper bound to the energy
eigenstates. This inconsistency can spoil the physical equivalence
of the two theories at high energies, but it is not altogether unexpected.
As the energy grows, that is to say, as $n$ grows, the wave functions
spread farther out in space and become highly non-local. Any canonical
transformation, on the contrary, is only locally valid, and thus our
time-dependent dilation transformation $\hat{D}$ (\ref{eq:time-dep-unitary-transf})
cannot guarantee there are no global issues in establishing the physical
equivalence of the two theories (or any two theories, for that matter).
This is not unexpected, since in general the non-local properties
of the wave functions can spoil any equivalence that the two theories
might have on the level of the Heisenberg equations. 

Darboux's method for solving the inverse problem of the calculus of
variations for second-order equations of motion of one-dimensional
systems produces theories which, despite being classically equivalent,
do not possess the expected quantum properties. Thus, whereas every
classical one-dimensional system can be ``Lagrangianized'', in quantum
mechanics one still encounters one-dimensional systems which can still
be called ``non-Lagrangian''. For such systems, quantum anomalies
resulting from limiting processes (in our case, the limit $t\rightarrow\infty$)
can be identified already at the classical level by an asymptotic
inequivalence between the original second-order equation and the EL
equation resulting from the Lagrangianization process. 

We note that the quantum Hamiltonian (\ref{eq:kg-quantum-hamiltonian})
naturally appears in the master equation for open quantum systems
in the description of the damped harmonic oscillator in the Lindblad
theory \cite{Isar1999} and Dekker's complex symplectic approach \cite{DEKKER1979}.
This relation has yet to be clarified, and it should prove useful
in the comparison of the BCK theory and the first-order theory along
the lines of a subsystem plus reservoir approach.

\paragraph*{Acknowledgements: M.C. Baldiotti and D.M. Gitman thank FAPESP for
financial support.}

\appendix
%dummy comment inserted by tex2lyx to ensure that this paragraph is not empty

\section{Appendix}

\paragraph*{Large $n$ limit of $\hat{A}\psi_{n}^{BCK}$}

Let us compute the action of $\hat{A}=\hat{q}\hat{p}+\hat{p}\hat{q}$
on the functions $\psi_{n}^{BCK}$ (\ref{eq:pseudostationary-state}),\begin{eqnarray}
\left(\hat{q}\hat{p}+\hat{p}\hat{q}\right)\psi_{n}^{BCK} & = & -i\left(2q\frac{\partial}{\partial q}+1\right)\psi_{n}^{BCK}\nonumber \\
 & = & i\sqrt{\left(n+2\right)\left(n+1\right)}\psi_{n+2}^{CK}-i\sqrt{n\left(n-1\right)}\psi_{n-2}^{BCK}\nonumber \\
 &  & -\frac{\alpha}{\tilde{\omega}}\left[\sqrt{\left(n+2\right)\left(n+1\right)}\psi_{n+2}^{CK}+\left(2n+1\right)\psi_{n}^{CK}+\sqrt{n\left(n-1\right)}\psi_{n-2}^{BCK}\right]\,.\label{eq:acalculus}\end{eqnarray}
 The asymptotic form of the pseudostationary states for large values
of $n$ is \cite{smirnov}\begin{eqnarray}
\psi_{2n}^{BCK}\left(q,t\right) & = & \left(\frac{\tilde{\omega}}{\pi}\right)^{1/4}\exp\left(-i\alpha\frac{q^{2}e^{2\alpha t}}{2}\right)\left(-1\right)^{n}\sqrt{\frac{\left(2n-1\right)!!}{2n!!}}\cos\left(\sqrt{\left(4n+1\right)\tilde{\omega}}e^{\alpha t}q\right)+O\left(n^{-1/4}\right)\nonumber \\
 & = & \psi_{2n}^{asym}+O\left(n^{-1/4}\right)\,,\nonumber \\
\psi_{2n+1}^{BCK}\left(q,t\right) & = & \left(\frac{\tilde{\omega}}{\pi}\right)^{1/4}\exp\left(-i\alpha\frac{q^{2}e^{2\alpha t}}{2}\right)\left(-1\right)^{n}\sqrt{\frac{\left(2n-1\right)!!}{2n!!}}\sin\left(\sqrt{\left(4n+3\right)\tilde{\omega}}e^{\alpha t}q\right)+O\left(n^{-1/4}\right)\nonumber \\
 & = & \psi_{2n+1}^{asym}+O\left(n^{-1/4}\right)\,.\label{eq:asymptotic-waves}\end{eqnarray}
Taking into account that $\cos\sqrt{4n+\alpha}x-\cos\sqrt{4n+1}x=O\left(n^{-1/4}\right)$
for any real $\alpha$, and substituting the above in the right-hand
side of (\ref{eq:acalculus}), one has\[
\left(\hat{q}\hat{p}+\hat{p}\hat{q}\right)\psi_{n}^{BCK}=-i\psi_{n}^{asym}+O\left(n^{-1/4}\right)\,.\]
On the other hand $\left(\hat{q}\hat{p}+\hat{p}\hat{q}\right)\psi_{n}^{BCK}=\left(\hat{q}\hat{p}+\hat{p}\hat{q}\right)\psi_{n}^{asym}+O\left(n^{-1/4}\right)$,
so\[
\left(\hat{q}\hat{p}+\hat{p}\hat{q}\right)\psi_{n}^{asym}=-i\psi_{n}^{asym}+O\left(n^{-1/4}\right)\,,\]
or to the same approximation,\[
\left(\hat{q}\hat{p}+\hat{p}\hat{q}\right)\psi_{n}^{BCK}=-i\psi_{n}^{BCK}+O\left(n^{-1/4}\right)\,.\]

\paragraph*{Proposition: The asymptotic pseudostationary functions $\psi_{n}^{asym}$
are not in the domain of $\hat{A}=\hat{q}\hat{p}+\hat{p}\hat{q}$.}

We first find the domain of $\hat{A}$ for which it is symmetric.
Consider, for $\phi,\psi\in D\left(\hat{A}\right)$:\begin{eqnarray*}
\left\langle \phi,\hat{A}\psi\right\rangle -\left\langle \hat{A}\phi,\psi\right\rangle  & = & -2i\int_{-\infty}^{\infty}dq\frac{d}{dq}\left(q\bar{\phi}\left(q\right)\psi\left(q\right)\right)\\
 & = & -2i\lim_{q\rightarrow\infty}q\left(\bar{\phi}\left(q\right)\psi\left(q\right)+\bar{\phi}\left(-q\right)\psi\left(-q\right)\right)\,.\end{eqnarray*}
Therefore, functions such that $\lim_{x\rightarrow\pm\infty}q\psi\left(q\right)\neq0$
are not in the domains of $\hat{A}$. On the other hand, we know that
for large values of $n$ the pseudostationary functions have the asymptotic
form (\ref{eq:asymptotic-waves}). Thus, clearly $\lim_{q\rightarrow\pm\infty}q\psi_{n}^{asym}\left(q\right)\ne0$
and $\psi_{n}^{asym}\neq D\left(\hat{A}\right)$. Note that the closure
of $\hat{A}$ is not affected by the exclusion of the asymptotic functions,
since they can't be the limit of any sequence.

\paragraph*{Proposition: $\hat{A}$ is self-adjoint.}

It suffices to show that the equation $\hat{A}^{*}\psi=\pm i\psi$
does not have solutions in $D\left(\hat{A}^{*}\right)$ \cite{reedsimonv1}.
The solutions are of the form\[
\psi_{\pm}=x^{\lambda_{\pm}}\,,\,\,\lambda_{\pm}=\mp\frac{1}{2}\,,\]
which are not square-integrable in the interval $\left(-\infty,\infty\right)$.
Therefore, the corresponding deficiency indices are $\left(0,0\right)$
and $\hat{A}$ is essentially self-adjoint.

\bibliographystyle{unsrt}
\bibliography{/home/fresneda/work/mendeley.bib/library}

\end{document}